\documentclass[a4paper,leqno]{CSCsimai}
\newcommand{\DOI}{DOI: 10.1685/CSC09XXX}
\newcommand{\newatop}[2]{\genfrac{}{}{0pt}{}{#1}{#2}}
\newcommand{\zero}{{\mathbf{0}}}
\newcommand{\puno}{{\mathbf{u}}}
\newcommand{\muno}{{\mathbf{d}}}
\newcommand{\eC}{\mathbf{c}^\textrm{e}}
\newcommand{\oC}{\mathbf{c}^\textrm{o}}
\newcommand{\cC}{\mathbf{c}}
\newcommand{\cF}{\mathbf{f}}

\begin{document}
\title{Competitive nucleation in metastable systems}
\author{{Emilio N.M.\ Cirillo$^1$, Francesca R.\ Nardi$^{2,3}$, 
         Cristian Spitoni$^4$}}
\address{$^1$
Dipartimento Me.Mo.Mat.,
Universit\`a degli Studi di Roma ``La Sapienza",\\
via A.\ Scarpa 16, I--00161, Roma, Italy\\
cirillo@dmmm.uniroma1.it}
\address{$^2$
Department of Mathematics,
Eindhoven University of Technology,\\
P.O.\ Box 513, 5600 MB Eindhoven, The Netherlands\\
F.R.Nardi@tue.nl}
\address{$^3$
Eurandom, P.O.\ Box 513, 5600 MB, Eindhoven, The Netherlands}
\address{$^4$
Department of Medical Statistics and Bioinformatics,
Leiden University Medical Centre,\\
Postal Zone S-05-P, PO Box 9600, 2300 RC Leiden The Netherlands\\
C.Spitoni@lumc.nl}

\begin{abstract}
Metastability is observed when a 
physical system is close to a first order phase transition.
In this paper the metastable behavior of a two state reversible 
probabilistic cellular automaton with self--interaction 
is discussed. Depending on the self--interaction, competing metastable 
states arise and a behavior very similar to that of the three 
state Blume--Capel spin model is found.
\end{abstract}

\keywords{Metastability, phase transition, 
          cellular automata.}

\section{Introduction.}
\label{intro}

\begin{figure}[htp]
\center\epsfig{file=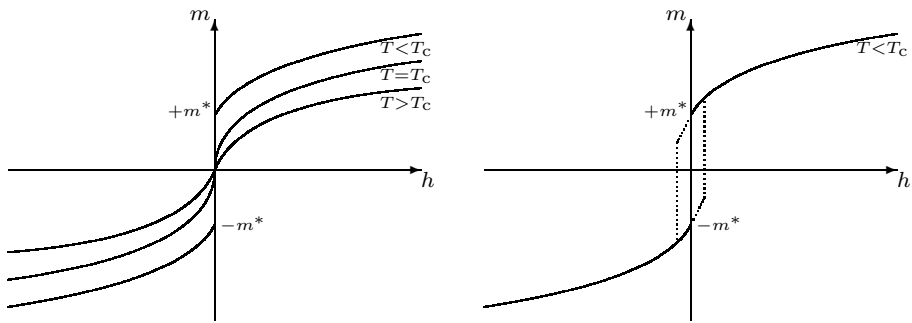,width=12.cm}
\caption{Isotherms of a ferromagnet on the left; ferromagnetic 
         hysteresis on the right. The temperature is denoted by $T$, 
         $m$ is the magnetization, $h$ is the external magnetic field,
         $m^*$ is the spontaneous magnetization, $T_\textrm{c}$ is the 
         Curie temperature.}
\label{fig:isteresi}
\end{figure}

Metastable states are observed when a 
physical system is close to a first order phase transition. Well known
examples
are super-saturated vapor states and magnetic hystereses~\cite{[OV]}.
In the Figure~\ref{fig:isteresi} the isotherms of a ferromagnet are 
depicted on the left; $T$ denotes the temperature,
$m$ the magnetization, i.e., the density of total magnetic moment,
$h$ the external magnetic field, $m^*>0$ the spontaneous magnetization, 
and $T_\textrm{c}$ the Curie temperature.
At temperature higher than $T_\textrm{c}$ 
the magnetization is zero for $h=0$; it is said 
that the system is in the \textit{paramagnetic} phase~\cite{huang}.
Below the critical temperature at $h=0$ the system
can exhibit the not zero values $m^*$ and $-m^*$; it is said that 
the system is in the \textit{ferromagnetic} phase. 
For $h=0$, when the temperature reaches the 
\textit{critical} value $T_\textrm{c}$ the system 
undergoes a \textit{continuous} (\textit{second order}) phase transition; 
the name is justified since the \textit{order parameter} $m$ varies
continuously when $T_\textrm{c}$ is crossed.

In the graph on the right in Figure~\ref{fig:isteresi} the behavior of the 
ferromagnet at $T$ smaller than $T_\textrm{c}$ is  
illustrated. When $h=0$ the system jumps from the positive magnetization
phase to the negative magnetization one, or vice-versa; the transition is 
called \textit{first order} since the order parameter $m$, which is 
the first derivative of one of the thermodynamical potential,
undergoes an abrupt variation~\cite{huang}.
Sometimes, provided the value $h=0$ is crossed sweetly in the experiment, 
the system persists in the same phase and the \textit{hysteresis} 
in the picture is observed. It is then said that the phase 
with negative (resp.\ positive) magnetization is \textit{metastable}
for $T<T_\textrm{c}$ and $h>0$ (resp.\ $h<0$) small.

The rigorous mathematical 
description of this phenomenon is relatively recent.
Not completely rigorous approaches based on equilibrium states have been 
developed in different fashions.
The purely dynamical point of view revealed more powerful and 
leaded to a pretty elegant definition and characterization of 
the metastable states; the most important results in this
respect have been summed up in~\cite{[OV]}.

In this paper we stick to the dynamical description 
and investigate competing metastable states. 
This problem shows up in connection with many physical processes,
such as the crystallization of proteins \cite{[tWF]}
and in glasses, in which the presence of a huge number 
of minima of the energy landscape prevents the system from reaching the 
equilibrium \cite{[BK]}.
The study of these systems is difficult, since 
the minima of the energy and the decay pathways between them 
change when the control parameters are varied. 
It is then of interest the study of models in which  
a complete control of the variations induced on the energy landscape by 
changes in the parameters is possible.
In Section~\ref{blume} we discuss the metastable behavior 
of the Blume--Capel model relying on results in~\cite{CO}.
In Section~\ref{pca} the obtained result will be compared with the 
known metastable behavior of reversible 
Probabilistic Cellular automata with self--interaction.

\section{The Blume--Capel model.}
\label{blume}
The Blume--Capel model has been introduced in~\cite{[B],[C]} in 
connection with the liquid Helium transition. 
In the context of metastability this model revealed very 
interesting for the three--fold nature of its ground states,
see~\cite{CO,[FGRN]}. 
Consider the two--dimensional torus $\Lambda=\{0,\dots,L-1\}^2$,
with $L$ even, endowed with the Euclidean metric;
$x,y\in\Lambda$ are \textit{nearest neighbors} iff
their mutual distance is equal to $1$.
Associate a variable $\sigma(x)=0,\pm1$
with each site $x\in\Lambda$ and let 
$\Omega=\{-1,0,+1\}^{\Lambda}$ be the 
\textit{configuration space}.
The \textit{energy} associated to the configuration $\sigma\in\Omega$ is 
\begin{equation}
\label{ham-blume}
H(\sigma)=
\sum_{<x,y>}(\sigma(x)-\sigma(y))^{2}
-\lambda\sum_{x\in\Lambda}(\sigma(x))^{2}
-h\sum_{x\in\Lambda}\sigma(x)
\end{equation}
where $<x,y>$ denotes a generic pair of nearest
neighbors sites in the torus 
$\Lambda$, 
$\lambda\in\mathbb{R}$ is the \textit{chemical potential},
$h\in\mathbb{R}$ is the \textit{external magnetic field}, and
$|h|,|\lambda|<1$.
The function $H$ will be also called \textit{Hamiltonian}.
The equilibrium behavior of the system is described by 
the \textit{Gibbs measure}
$\mu(\sigma):=\exp\{-\beta H(\sigma)\}/Z$, where $\beta$ is the inverse 
of the temperature and the normalization constant $Z$ is called 
\textit{partition function}.

It is possible to introduce the stochastic version of the model 
by defining a serial dynamics reversible w.r.t.\ the Hamiltonian 
(\ref{ham-blume}). 
It will be a discrete time Glauber dynamics, that is a Markov chain with 
state space $\Omega$ and \textit{transition matrix} 
$p:\Omega\times\Omega\to[0,1]$ such that
\begin{equation}
\label{metro01}
p(\sigma ,\eta):=
\frac{1}{2|\Lambda|} e^{-\beta \max\{H(\eta)-H(\sigma),0\}}
\end{equation}
for $\sigma,\eta\in\Omega$ such 
$\sigma$ and $\eta$ are \textit{nearest neighboring configurations}, i.e.,
$\sigma$ is equal to $\eta$ excepted 
for the value of the spin associated to a single site;  
$p(\sigma,\eta):=0$ 
for $\sigma,\eta\in\Omega$ such that $\sigma\neq\eta$ and $\sigma$ 
and $\eta$ are not nearest neighboring, that is to say they differ for the 
values of the spins associated to at least two sites.
To ensure the correct normalization of the transition matrix,
we also set $p(\sigma,\sigma)=1-\sum_{\eta\neq\sigma}p(\sigma,\eta)$ for 
any $\sigma\in\Omega$.

This dynamics, called \textit{Metropolis algorithm}, satisfies the two 
following important properties:
(i) only transitions between nearest neighboring configurations are allowed;
(ii) the dynamics is \textit{reversible} w.r.t.\ the Hamiltonian 
(\ref{ham-blume}), i.e., 
\begin{equation}
\label{dettagliato}
\mu(\sigma)p(\sigma,\eta)=\mu(\eta)p(\eta,\sigma)
\end{equation}
for any $\sigma,\eta\in\Omega$. The equation (\ref{dettagliato}) is 
called \textit{detailed balance} condition.

The definition (\ref{metro01}) of the dynamics implies that 
transitions decreasing the energy happen with finite probability,"
while transitions increasing the energy are performed with probability 
tending to zero for $\beta\to\infty$, that is when the temperature 
tends to zero. This means that when the temperature is small, the system
takes a time exponentially large in $\beta$ to leave 
a \textit{local minimum} of the Hamiltonian, 
i.e., a configuration $\sigma\in\Omega$ such that 
$H(\eta)>H(\sigma)$ for any $\eta\in\Omega$ nearest neighbor of 
$\sigma$.
We can then expect that, wherever started, the systems tends to reach 
the \textit{ground state} of the energy, i.e., the minimum of $H$,
in a \textit{tunneling time} depending on the initial condition.  
Supposing that there exists initial data for which the tunneling time
is exponentially large in $\beta$, it is rather 
natural to define the metastable state as the configuration to which 
corresponds the maximum tunneling time. 

\begin{figure}[htp]
\center\epsfig{file=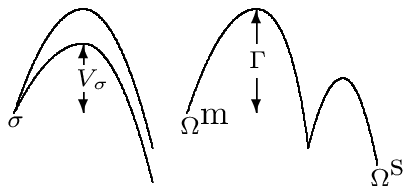,width=6.cm}
\caption{Definition of metastable states.}
\label{fig:definizione}
\end{figure}

More precisely, 
following~\cite{[OV]} and referring to the Figure~\ref{fig:definizione}
for a description of the following definitions, 
given a sequence of configurations
$\omega=\omega_1,\dots,\omega_n$, with $n\ge2$, we define the 
\textit{energy height} along the path $\omega$ as  
$\Phi_\omega=\max_{i=1,\dots,|\omega|} H(\omega_i)$.
Given $A,A'\subset\Omega$,
we let the \textit{communication energy} $\Phi(A,A')$ 
between $A$ and $A'$ be 
the minimal energy height $\Phi_\omega$ over the set of paths $\omega$
starting in $A$ and ending in $A'$.
For any $\sigma\in\Omega$, we let
$\mathcal{I}_\sigma\subset\Omega$
be the set of configurations with energy strictly below $H(\sigma)$ and
$V_\sigma=\Phi(\sigma,\mathcal{I}_\sigma)-H(\sigma)$ 
be the \textit{stability level
of} $\sigma$, that is the energy barrier that, starting from $\sigma$, must be 
overcome to reach the set of configurations with energy smaller than 
$H(\sigma)$; we set $V_\sigma=\infty$ if $\mathcal{I}_\sigma=\emptyset$.
We denote by $\Omega^\textrm{s}$ the set of
global minima of the energy (\ref{ham-blume}), i.e., the collection of the 
ground states, and suppose that 
the \textit{communication energy}
$\Gamma=\max_{\sigma\in\Omega\setminus\Omega^\textrm{s}}V_\sigma$
is strictly positive.
Finally, we define the set of \textit{metastable states}
$\Omega^\textrm{m}=\{\eta\in\Omega:\,V_\eta=\Gamma\}$.
The set $\Omega^\textrm{m}$ deserves its name, 
since in a rather general framework it is possible to prove 
(see, e.g., \cite[Theorem~4.9]{[MNOS]}) the following:
pick $\sigma\in\Omega^\textrm{m}$, consider the 
chain $\sigma_n$ started at 
$\sigma_0=\sigma$, then the \textit{first hitting time}
$\tau_{\Omega^\textrm{s}}=\inf\{t>0:\,\sigma_t\in\Omega^\textrm{s}\}$
to the ground states is a random variable with mean 
exponentially large in $\beta$, that is 
\begin{equation}
\label{mnos4.9}
\lim_{\beta\to\infty}
\frac{1}{\beta}\,\log\mathbb{E}_\sigma[\tau_{\Omega^\textrm{s}}]=\Gamma
\end{equation}
with $\mathbb{E}_\sigma$ the average on the trajectories 
started at $\sigma$.
In the considered regime, finite volume and temperature tending to zero, 
the description of metastability is then reduced
to the computation of $\Omega^\textrm{s}$, $\Gamma$, and 
$\Omega^\textrm{m}$.

After this rather general discussion on the definition of metastable states
we get back to the study of the Blume--Capel model and note that 
rigorous results have already been found in~\cite{CO} in the region 
$h>\lambda>0$. In this section we review those results on heuristic 
grounds and extend the discussion to the whole region 
$h>0$ and $h>-\lambda$. 

\begin{figure}[htp]
\center\epsfig{file=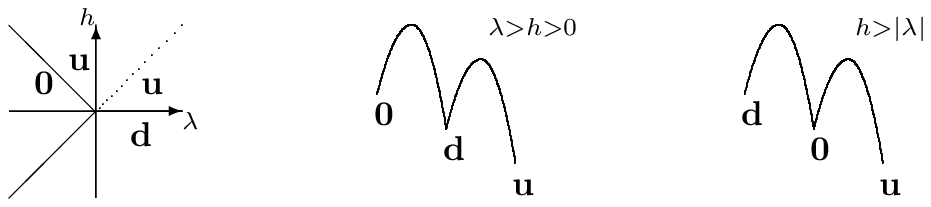,width=12.cm}
\caption{Ground states of the Blume--Capel model.}
\label{fig:blume}
\end{figure}

First of all we describe the structure of the ground states of the 
Hamiltonian.
Denote by $\muno$, $\puno$ and $\zero$
the configurations with all 
the spins in $\Lambda$ equal respectively to $-1$, $+1$ and $0$, 
and remark that 
$E(\puno)=-L^2(\lambda+h)$, 
$E(\muno)=-L^2(\lambda-h)$, 
and
$E(\zero)=0$.
It is not difficult to prove that 
for $\lambda=h=0$ the ground state is three times degenerate and the 
configurations minimizing the Hamiltonian are $\muno$, $\puno$ 
and $\zero$;
for $h>0$ and $h>-\lambda$, the ground state is $\puno$;
for $h<0$ and $h<\lambda$ the ground state is $\muno$;
for $\lambda<0$ and  $\lambda<h<-\lambda$ the ground state is $\zero$;
for $h=0,\lambda>0$ the ground state is two times degenerate and the 
configurations minimizing the Hamiltonian are $\muno$ and $\puno$; 
for $h=\lambda<0$ the ground state is two times degenerate and the 
configurations minimizing the Hamiltonian are $\muno$ and $\zero$; 
for $h=-\lambda>0$ the ground state is two times degenerate and the 
configurations minimizing the Hamiltonian are $\puno$ and $\zero$.
These results are summarized in the graph in the left in 
Figure~\ref{fig:blume}.
Note, also, that 
$E(\zero)>E(\muno)>E(\puno)$ for $0<h<\lambda\le1$, 
$E(\zero)=E(\muno)>E(\puno)$ for $0<h=\lambda\le1$, 
and
$E(\muno)>E(\zero)>E(\puno)$ for $h>|\lambda|$,
see the two graphs on the right in the Figure~\ref{fig:blume}.

The obvious candidates to be metastable states 
are the configurations $\mathbf{d}$ or $\mathbf{0}$; in particular the 
situation in the region $h>\lambda>0$ looks really intriguing.
In order to prove 
rigorously that one of them is the metastable state, one should 
compute $\Gamma$ and prove that either $V_\muno$ or 
$V_\zero$ is equal to $\Gamma$. 
This is a difficult task, indeed all the paths $\omega$ 
connecting $\muno$ and $\zero$ to $\puno$ should be taken 
into account and the related energy heights $\Phi_\omega$ computed.
This problem has been solved rigorously in~\cite{CO} in the region 
$h>\lambda>0$ under the technical 
restriction $2(h/\lambda)^2+h/\lambda-1<2J/\lambda$.
There it has been proven that the metastable state is $\muno$ and 
that, depending on the ration $h/\lambda$, during the tunneling from the 
metastable to the stable state the configuration $\zero$ is visited 
or not visited.

As mentioned above we develop an heuristic argument to characterize 
the behavior of the system in the whole region 
$h>0$ and $h>-\lambda$. 
To characterize the local minima of the Hamiltonian, it is necessary 
to compute the energy variation under the flip of a single spin. Then consider
$\sigma\in\Omega$, $x\in\Lambda$, $a\in\{-1,0,+1\}$, and denote 
by $\sigma_x^a$ the configuration such that 
$\sigma_x^a(y)=\sigma(y)$ for all $y\neq x$ and $\sigma_x^a(x)=a$; note that 
$\sigma_x^a=\sigma$ iff $a=\sigma(x)$.
By using (\ref{ham-blume}) we easily get
\begin{equation}
\label{delta}
H(\sigma_x^a)-H(\sigma)
=-2
 (a-\sigma(x))
 S_\sigma(x)
-(\lambda-4)
 (a^2-\sigma(x)^2)
-h
 (a-\sigma(x))
\end{equation}
where $S_\sigma(x)$ is the sum of the four spins of $\sigma$ associated 
to the nearest neighbors of the site $x$.
Equation (\ref{delta}) can be used to compute the energy difference 
involved in all the possible spin flips; the results are 
summarized in the Table~\ref{tab:blume}.
Note that the three cases not listed in the table can be deduced by 
changing the sign accordingly, for instance if 
$\sigma(x)=-1$ and $a=+1$, we get 
$H(\sigma_x^a)-H(\sigma)=-4S_\sigma(x)-2h$ whose sign is 
positive for $S_\sigma(x)\le-1$ and negative for $S_\sigma(x)\ge0$.
It is also worth remarking that the results on the sign 
of the energy differences listed in the third column of the 
Table~\ref{tab:blume} strongly depend on the assumption 
$|\lambda|,|h|<1$.

\begin{table}[htp]
\caption{Spin flip energy costs. In the last column the sign of the 
         energy difference is discussed.}
\label{tab:blume}
\medskip
\begin{center}
    \leavevmode
    \begin{tabular}{c|c|c|c} 
\hline\hline
$\sigma(x)$ & $a$ & $H(\sigma_x^a)-H(\sigma)$ & sign \\
\hline\hline
$+1$ & $-1$ & $4S_\sigma(x)+2h$ &
\begin{tabular}{l}
$>0$ if $S_\sigma(x)\ge0$\\
$<0$ if $S_\sigma(x)\le-1$\\
\end{tabular}
\\
\hline
$+1$ & $0$ & $2S_\sigma(x)-4+\lambda+h$ &
\begin{tabular}{l}
$>0$ if $S_\sigma(x)\ge+2$\\
$<0$ if $S_\sigma(x)\le+1$\\
\end{tabular}
\\
\hline
$0$ & $-1$ & $2S_\sigma(x)+4-\lambda+h$ &
\begin{tabular}{l}
$>0$ if $S_\sigma(x)\ge-1$\\
$>0$ if $S_\sigma(x)\ge-2$ and $h>\lambda$\\
$<0$ if $S_\sigma(x)\le-2$ and $h<\lambda$\\
$<0$ if $S_\sigma(x)\le-3$\\
\end{tabular}
\\
\hline\hline
    \end{tabular}
\end{center}
\end{table}

From the results in Table~\ref{tab:blume} it follows that 
for $h>\lambda$ 
the local configurations in which a minus can appear in a local
minimum are those such that the sum of the neighboring spins 
is smaller than or equal to $-3$, see the two configurations 
on the left in Figure~\ref{f:loc02}. 
For $h<\lambda$ 
the local configurations in which a minus can appear in a local
minimum are those such that the sum of the neighboring spins 
is smaller than or equal to $-2$, see the four configurations in 
the Figure~\ref{f:loc02}. 
\begin{figure}[htp]
\hskip 1.5 cm
\begin{minipage}{1.5 cm}
\begin{center}
$-$\\
$-$ $-$ $-$\\
$-$
\end{center}
\end{minipage}
\hskip 1 cm
\begin{minipage}{1.5 cm}
\begin{center}
$0$\\
$-$ $-$ $-$\\
$-$
\end{center}
\end{minipage}
\hskip 1 cm
\begin{minipage}{1.5 cm}
\begin{center}
$0$\\
$0$ $-$ $-$\\
$-$
\end{center}
\end{minipage}
\hskip 1 cm
\begin{minipage}{1.5 cm}
\begin{center}
$+$\\
$-$ $-$ $-$\\
$-$
\end{center}
\end{minipage}
\vskip 0.1 cm
\caption{Minus spins allowed in a local minimum; for $h>\lambda$
         only the two configurations on the left are allowed, 
         while for $h<\lambda$ all the four depicted configurations are 
         possible.}
\label{f:loc02}
\end{figure}

From the first two lines in Table~\ref{tab:blume} it follows that 
the sole local configurations in which a plus spin can appear in a local
minimum are those such that the sum of the neighboring spins 
is greater than or equal to $2$, see Figure~\ref{f:loc01}.

\begin{figure}[htp]
\hskip 1.5 cm
\begin{minipage}{1.5 cm}
\begin{center}
$+$\\
$+$ $+$ $+$\\
$+$
\end{center}
\end{minipage}
\hskip 1 cm
\begin{minipage}{1.5 cm}
\begin{center}
$0$\\
$+$ $+$ $+$\\
$+$
\end{center}
\end{minipage}
\hskip 1 cm
\begin{minipage}{1.5 cm}
\begin{center}
$-$\\
$+$ $+$ $+$\\
$+$
\end{center}
\end{minipage}
\hskip 1 cm
\begin{minipage}{1.5 cm}
\begin{center}
$0$\\
$0$ $+$ $+$\\
$+$
\end{center}
\end{minipage}
\vskip 0.1 cm
\caption{Plus spins allowed in a local minimum.}
\label{f:loc01}
\end{figure}

\begin{figure}[htp]
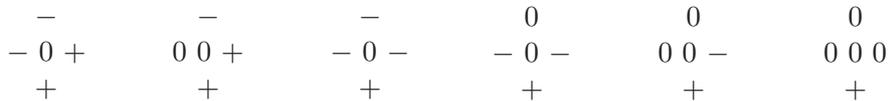

\begin{minipage}{1.5 cm}
\begin{center}
$-$\\
$-$ $0$ $+$\\
$+$
\end{center}
\end{minipage}
\hskip 0.5 cm
\begin{minipage}{1.5 cm}
\begin{center}
$-$\\
$0$ $0$ $+$\\
$+$
\end{center}
\end{minipage}
\hskip 0.5 cm
\begin{minipage}{1.5 cm}
\begin{center}
$-$\\
$-$ $0$ $-$\\
$+$
\end{center}
\end{minipage}
\hskip 0.5 cm
\begin{minipage}{1.5 cm}
\begin{center}
$0$\\
$-$ $0$ $-$\\
$+$
\end{center}
\end{minipage}
\hskip 0.5 cm
\begin{minipage}{1.5 cm}
\begin{center}
$0$\\
$0$ $0$ $-$\\
$+$
\end{center}
\end{minipage}
\hskip 0.5 cm
\begin{minipage}{1.5 cm}
\begin{center}
$0$\\
$0$ $0$ $0$\\
$+$
\end{center}
\end{minipage}
\vskip 0.1 cm
\caption{Zero spins with at least a neighboring plus 
         allowed in a local minimum for $h>\lambda$.}
\label{f:loc04}
\end{figure}

\begin{figure}[htp]
\center\epsfig{file=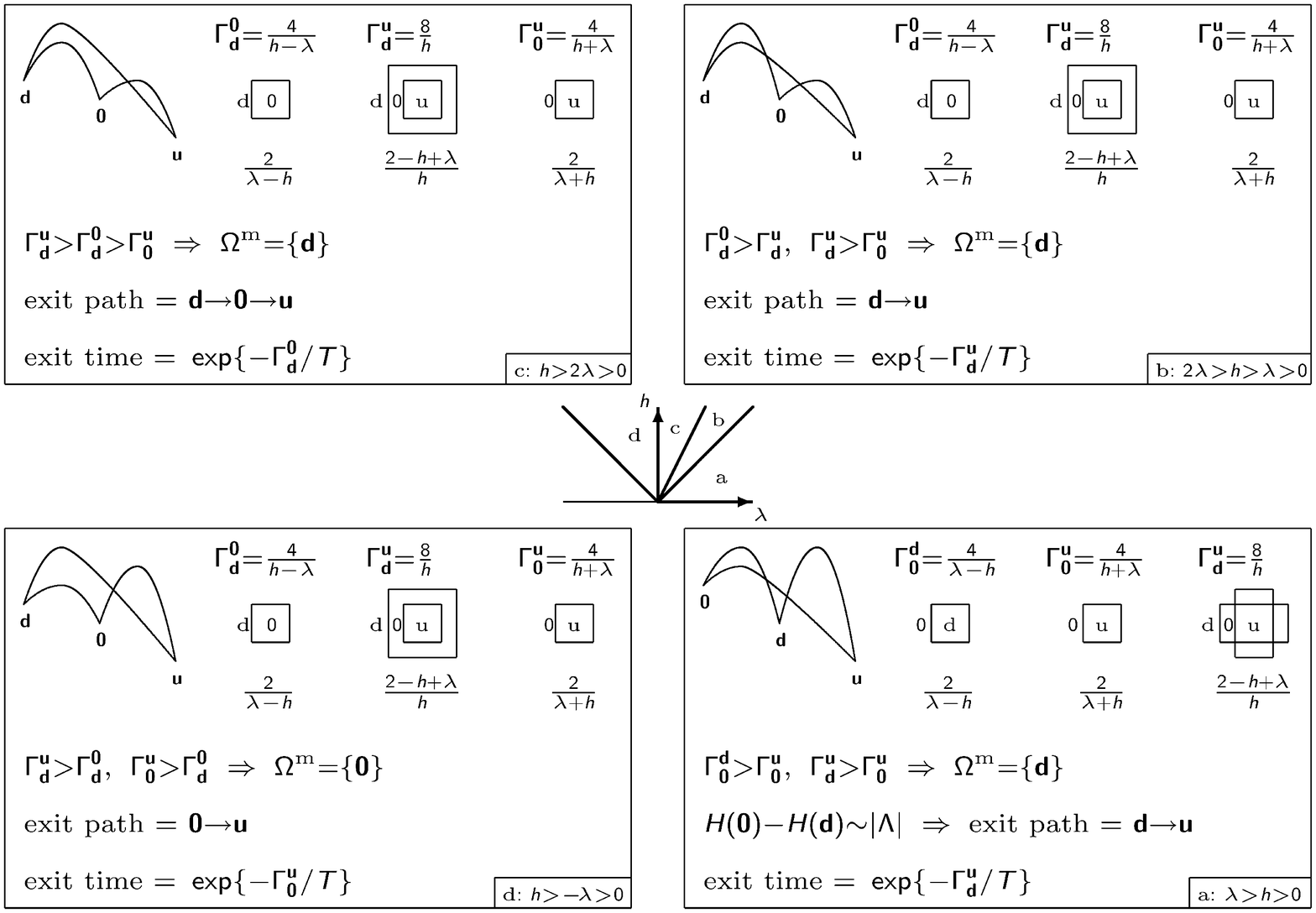,width=12.cm}
\caption{Summary of results for the Blume--Capel model.}
\label{fig:blume2}
\end{figure}

We discuss in detail the case $h>\lambda$; the analogous results in the 
region $\lambda>h>0$ will be summarized in the 
Figure~\ref{fig:blume2}. From the necessary condition 
for a minus in a local minimum, see 
the two graphs on the left in the Figure~\ref{f:loc02}, we have that 
for a configuration to be a local minimum it is necessary that 
the zeroes form well separated rectangles possibly winding around the torus.
To verify that this condition is sufficient for the configuration to be 
a local minimum we note that, in this case $h>\lambda$,
the local configurations in which a zero can appear in a local
minimum are those such that the sum of the neighboring spins 
is greater than or equal to $-2$ and smaller than or equal to $+1$.
In the Figure~\ref{f:loc04} the possible local configuration for a zero with at
least a neighboring plus are shown.
This condition is surely met in a configuration in which the zeroes 
form separated rectangular clusters plunged in a sea of minuses 
with side lengths larger or equal to two. 
Moreover, see the Figure~\ref{f:loc02}, in a local minimum direct interfaces
between minuses and pluses are forbidden, then the pluses must necessarily 
be located in the bulk of the zero rectangular droplets. 
From the results in the Figure~\ref{f:loc04}, see in particular the 
two graphs on the right, it follows that the pluses must a form well 
separated rectangular clusters, possibly winding around the torus, inside 
a rectangular zero cluster. Note that the plus cluster can be separated 
by the minus component even by a single layer of zeroes.

In order to study the nucleation of the stable state starting from 
the possibly metastable states $\zero$ and $\muno$ the 
interesting local minima, in the case $h>\lambda$, are 
the \textit{zero rectangular droplets} in the see of minuses, 
the \textit{plus rectangular droplets} in the sea of zeroes, and 
the \textit{frames} made of a plus rectangular droplet plunged 
in the sea of minuses and separated by the minus component by a 
single layer of pluses (the frame).
The local minima can be used to construct 
the optimal paths connecting $\muno$ and $\zero$ to the ground state $\puno$.

Consider, first, the paths from $\muno$ to $\zero$.
Optimal paths can be reasonably constructed via a sequence of 
zero droplets.
The difference of energy between two zero droplets
with side lengths respectively given by $\ell,m\ge2$ and  $\ell,m+1$ is  
equal to $2-(h-\lambda)\ell$.
It then follows that the energy of a such a droplet is increased by 
adding an $\ell$--long slice iff 
$\ell<\lfloor2/(h-\lambda)\rfloor+1=\ell^\zero_\muno$, where
$\lfloor x\rfloor$ denotes the largest integer smaller than the 
real $x$.
The length $\ell^\zero_\muno$ is called the 
\textit{critical length}.
It is reasonable that the energy barrier $V_\zero$ is given
by the difference of energy between 
the smallest supercritical zero droplet,
i.e., the square zero droplet with side length 
$\ell^\zero_\muno$, and the configuration $\zero$;
by using (\ref{ham-blume}) we get that
such a difference of energy 
is equal to
$\Gamma^\zero_\muno=4/(h-\lambda)$.

A path from $\zero$ to $\puno$ can be constructed with a sequence of 
plus droplets.
By using (\ref{ham-blume}) we get that the difference of energy 
between two plus droplets with side lengths respectively given 
by $\ell,m\ge2$ and  $\ell,m+1$ is  equal to $2-2(h+\lambda)\ell$.
It then follows that the energy of a plus droplet is increased by 
adding an $\ell$--long slice iff 
$\ell<\lfloor2/(h+\lambda)\rfloor+1=\ell^\puno_\zero$.
The length $\ell^\puno_\zero$ is the 
critical length for the plus droplets;
the difference of energy between the smallest supercritical 
plus droplet and $\zero$ is equal to 
$\Gamma^\puno_\zero=4/(h+\lambda)$.

A path from $\muno$ to $\puno$ can be constructed via a 
sequence of frames.
It is not difficult to prove that 
the difference of energy 
between two frames with internal (rectangle of pluses) side lengths 
respectively given by $\ell,m\ge2$ and  $\ell,m+1$ is equal to 
$4-2(h-\lambda)-2h\ell$, so that the critical length for those frames 
is given by 
$\ell^\cF_\muno=\lfloor(2-(h-\lambda))/h\rfloor+1$ and 
the difference of energy between the smallest supercritical frame 
and $\muno$ is equal 
to 
$\Gamma^\cF_\muno=8+2(\ell^\cF_\muno)^2h-4h\ell^\cF_\muno\varepsilon
 -4(h-\lambda)$, where $\varepsilon=\ell^\cF_\muno-[(2-(h-\lambda)/h]$.

Remarked that for $h,\lambda\ll1$ one has 
$\Gamma^\cF_\muno\sim8/h$, by comparing the energy barriers computed above,
it is possible to find the communication energy $\Gamma$
and to deduce all the results summarized in the Figure~\ref{fig:blume2}.

\section{Probabilistic cellular automata with self--interaction.}
\label{pca}
We have seen above how in the case of a three--state model 
as the Blume--Capel model competing metastable states shows up. 
In some sense this result is natural because the single site configuration
space is 
three--state. In the framework of Probabilistic Cellular Automata 
it has been shown, see~\cite{[BCLS],[CN],[CNS],[CNS2]}, how 
competing metastable states arise in the context of a genuine two--state model.

Consider the two--dimensional torus $\Lambda=\{0,\dots,L-1\}^2$,
with $L$ even, endowed with the Euclidean metric.
Associate a variable $\sigma(x)=\pm1$
with each site $x\in\Lambda$ and let $\Omega=\{-1,+1\}^{\Lambda}$ be the 
\textit{configuration space}.
Let $\beta>0$ and $\kappa,h\in[0,1]$.
Consider the Markov chain $\sigma_n$, with $n=0,1,\dots$,
on $\Omega$ with {\it transition matrix}
\begin{equation}
\label{markov}
p(\sigma,\eta)
=\prod_{x\in\Lambda}p_{x,\sigma}\left(\eta(x)\right)\;\;\;
\forall\sigma,\eta\in\Omega
\end{equation}
where, for $x\in\Lambda$ and $\sigma\in\Omega$,
$p_{x,\sigma}(\cdot)$ is the probability measure on $\{-1,+1\}$
defined as 
$p_{x,\sigma}(s)
 =1/[1+\exp\left\{-2\beta s(S_\sigma(x)+h)\right\}]$
with $s\in\{-1,+1\}$ and
$S_{\sigma}(x)=\sum_{y\in\Lambda}K(x-y)\,\sigma(y)$
where 
$K(x-y)$ is $0$ if $|x-y|\ge2$,
$1$ if $|x-y|=1$, and
$\kappa$ if $|x-y|=0$.
The probability $p_{x,\sigma}(s)$ for the spin $\sigma(x)$ to be equal to $s$
depends only on the values of the spins of $\sigma$ 
in the five site cross centered at $x$.
The metastable behavior of model (\ref{markov}) has been studied
in Ref.\ \cite{[CN]} for $\kappa=0$ and in Ref.\ \cite{[BCLS],[CNS]} for
$\kappa=1$.  

The Markov chain (\ref{markov}) is a
\textit{probabilistic cellular automata} (PCA);
the chain $\sigma_n$, with $n=0,1,\dots$,
updates all the spins simultaneously and independently at any time.
The chain is \textit{reversible} 
with respect to the Gibbs measure
$\mu(\sigma)=\exp\{-\beta H(\sigma)\}/Z$
with
$Z=\sum_{\eta\in\Omega}\exp\{-\beta H(\eta)\}$
and
\begin{equation}
\label{ham}
H(\sigma)=
-h\sum_{x\in\Lambda}\sigma(x)
-\frac{1}{\beta}\sum_{x\in\Lambda}\log\cosh\left[\beta
\left(
S_{\sigma}(x)+h\right)\right]
\end{equation}
that is \textit{detailed balance}
$p(\sigma,\eta)\,e^{-\beta H(\sigma)}=
 p(\eta,\sigma)\,e^{-\beta H(\eta)}$
holds and, hence, $\mu$ is stationary;
$1/\beta$ is called the {\it temperature} and $h$ the
{\it magnetic field}. 

Although the dynamics is reversible w.r.t.\ the Gibbs measure associated
to the Hamiltonian (\ref{ham}), the probability 
$p(\sigma,\eta)$ cannot be expressed in terms of  
$H(\sigma)-H(\eta)$, as usually happens for Glauber dynamics. 
Given $\sigma,\eta\in\Omega$, we define the \textit{energy cost} 
\begin{equation}
\label{defdelta}
\Delta(\sigma,\eta)=
 -\lim_{\beta\to\infty}\frac{\log p(\sigma,\eta)}{\beta}
 =
\!\!\!\!\!\!\!\!\!
\sum_{\newatop{x\in\Lambda:}{\eta(x)[S_\sigma(x)+h]<0}}
\!\!\!\!\!\!\!\!\!
2|S_\sigma(x)+h|
\end{equation}
Note that $\Delta(\sigma,\eta)\ge0$ and $\Delta(\sigma,\eta)$ is not
necessarily equal to $\Delta(\eta,\sigma)$;
it can be proven, see \cite[Section~2.6]{[CNS]}, that 
\begin{equation}
\label{cri01}
e^{-\beta\Delta(\sigma,\eta)-\beta\gamma(\beta)}
\le
p(\sigma,\eta)
\le
e^{-\beta\Delta(\sigma,\eta)+\beta\gamma(\beta)}
\end{equation}
with $\gamma(\beta)\to0$ in the zero temperature limit $\beta\to\infty$.
Hence, $\Delta$ can be 
interpreted as the cost of the transition from $\sigma$ to $\eta$
and plays the role that, in the context of Glauber dynamics, is 
played by the difference of energy.

In this context 
the ground states are those configurations on which the Gibbs
measure $\mu$ concentrates when
$\beta\to\infty$; hence, they can be defined as the
minima of the \textit{energy}
\begin{equation}
\label{hl}
E(\sigma)=
\lim_{\beta\to\infty}H(\sigma)
=
-h\sum_{x\in\Lambda}\sigma(x)
-\sum_{x\in\Lambda}|S_{\sigma}(x)+h|
\end{equation}
For $X\subset\Omega$, we set 
$E(X)=\min_{\sigma\in X}E(\sigma)$.
For $h>0$ the configuration $\puno$, with
$\puno(x)=+1$ for $x\in\Lambda$, is the unique ground state,
indeed each site contributes to the energy with $-h-(4+\kappa+h)$. 
For $h=0$, the ground states are the configurations such that all the sites
contribute to the sum (\ref{hl}) with $4+\kappa$. 
Hence, for $\kappa\in(0,1]$, the sole ground states are the configurations 
$\puno$ and $\muno$, with $\muno(x)=-1$ for $x\in\Lambda$.
For $\kappa=0$, the configurations $\eC,\oC\in\Omega$ such that 
$\eC(x)=(-1)^{x_1+x_2}$ and $\oC(x)=(-1)^{x_1+x_2+1}$
for $x=(x_1,x_2)\in\Lambda$ are ground states, as well. Notice that 
$\eC$ and $\oC$ are chessboard--like states with the pluses  
on the even and odd sub--lattices, respectively; we set $\cC=\{\eC,\oC\}$. 
Since the side length $L$ of the torus $\Lambda$ 
is even, then $E(\eC)=E(\oC)=E(\cC)$.
By studying those energies as a function of $\kappa$ and $h$, recalling
that periodic boundary conditions are considered,
we get 
$E(\puno)=-L^2(4+\kappa+2h)$, 
$E(\muno)=-L^2(4+\kappa-2h)$, 
and
$E(\cC)=-L^2(4-\kappa)$;
hence
$E(\cC)>E(\muno)>E(\puno)$ for $0<h<\kappa\le1$, 
$E(\cC)=E(\muno)>E(\puno)$ for $0<h=\kappa\le1$, 
and
$E(\muno)>E(\cC)>E(\puno)$ for $0<\kappa<h\le1$.
  
In~\cite{[CNS2]} the metastable behavior of this model has been studied 
with an heuristic argument very similar to the one developed in 
the Section~\ref{blume} to discuss the metastable behavior of the 
Blume--Capel model. For the details we refer the interested 
reader to the quoted paper, we just mention here, that quite 
surprisingly results very similar to the ones obtained in the 
framework of the Blume--Capel model are found, provided the different 
parameters are interpreted according to the correspondences in 
Table~\ref{t:corrisp}.

\begin{table}[htp]
\caption{Correspondence between the Blume--Capel model and the 
PCA.}
\label{t:corrisp}
\medskip
\begin{center}
    \leavevmode
    \begin{tabular}{l|c|c|c|c|c} 
\hline\hline
Blume--Capel & $\puno$ & $\muno$ & $\zero$ & $h$ & $\lambda$ \\
\hline
PCA & $\puno$ & $\muno$ & $\cC$ & $h/2$ & $\kappa/2$ \\
\hline\hline
    \end{tabular}
\end{center}
\end{table}

\begin{figure}[htp]
\hskip 0.5 cm
\epsfig{file=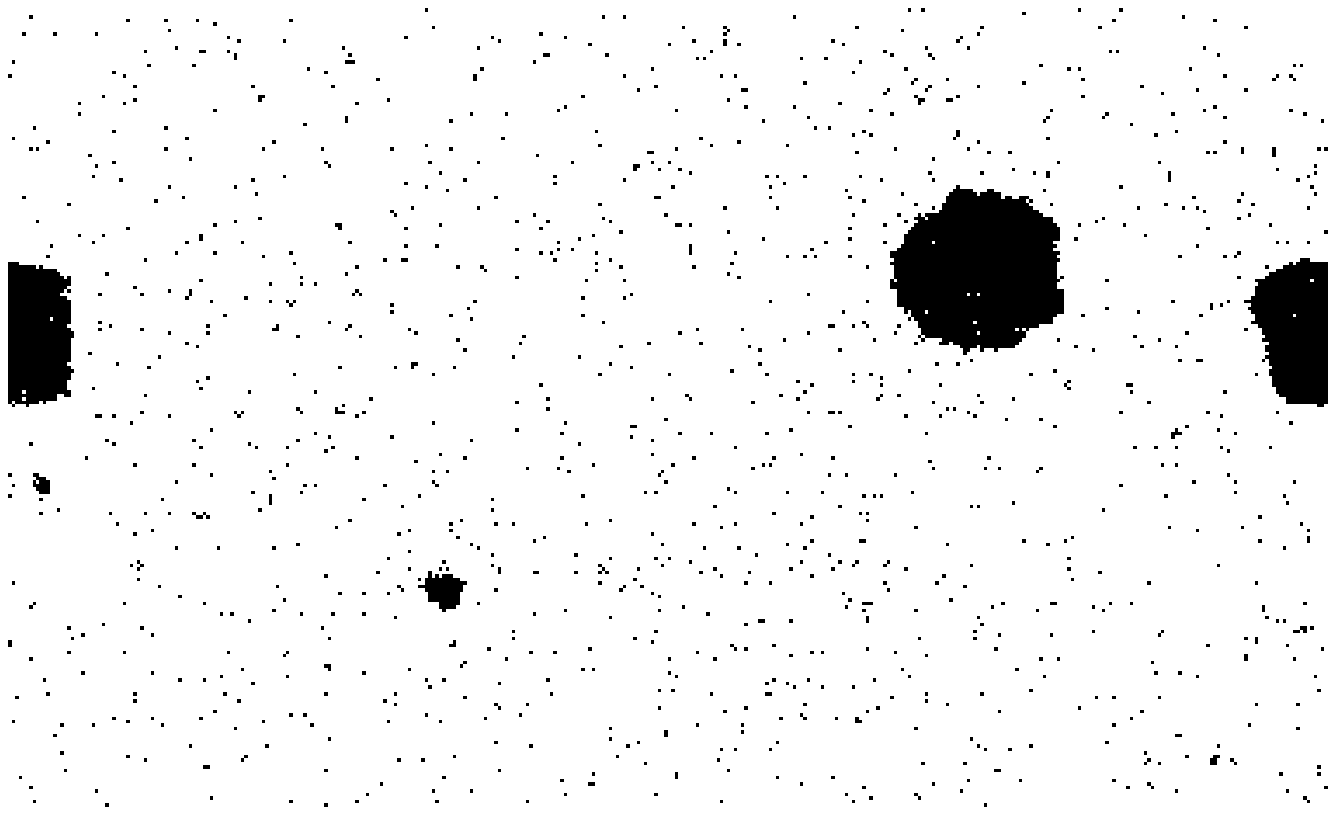,height=3.7cm,width=4.62cm}
\hskip 2 cm
\epsfig{file=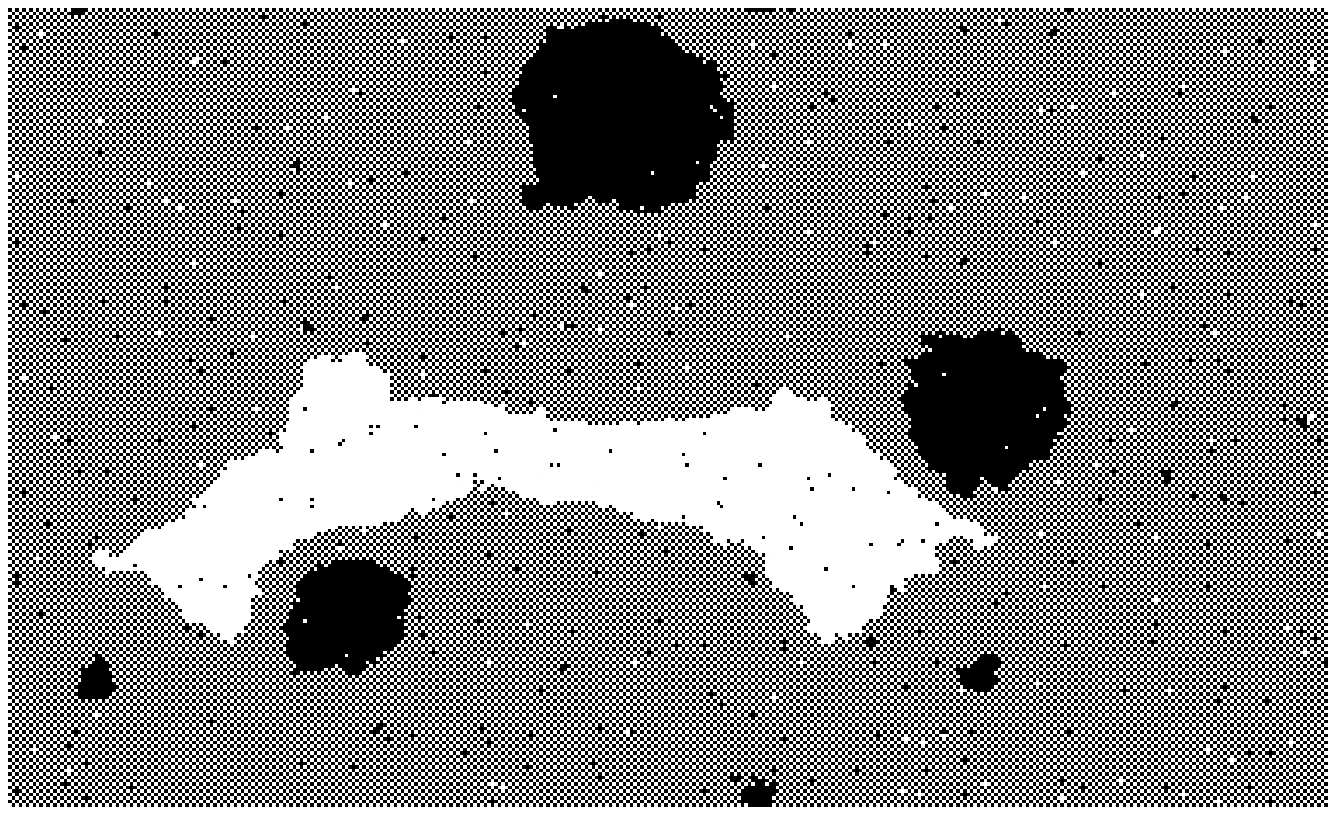,height=3.7cm,width=4.62cm}
\caption{On the left, typical configuration of the PCA 
         with $\kappa=1$; simulation performed 
         on a $380\times230$ torus with $\beta=0.7$. 
         White and black 
         points represent respectively minus and plus spins.
         On the right, the same with 
         $\kappa=0$; 
         the chessboard region looks grey.}
\label{fig:con}
\end{figure}

Notice that the role of the 
zero state of the Blume--Capel model is played, in the context of the 
PCA, by the flip--flopping
chessboard--like configurations.
As (\ref{mnos4.9}) shows, the discussed results are valid 
in the limit $\beta\to\infty$. Their validity at 
finite temperature can be tested with Monte Carlo simulations, see
the configurations in Figure~\ref{fig:con}
observed in a run of the dynamics 
of the PCA with the parameters 
specified in the caption and with starting configuration 
$\muno$. On the left it is shown that if the 
self--interaction is present the nucleation of the plus phase 
is achieved directly; the plot on the right shows that, if the 
self--interaction is zero, than the chessboard--like phase is visited 
before the plus phase is nucleated. 

\baselineskip=0.9\normalbaselineskip

\end{document}